\magnification=\magstep1
\baselineskip=17pt
\hfuzz=6pt

$ $

\vskip 1in
\centerline{\bf Hamiltonian singular value transformation and inverse block encoding}

\bigskip

\centerline{Seth Lloyd,$^{*1,2}$ 
Bobak T. Kiani,$^{1,3}$
David R.M. Arvidsson-Shukur$^{4}$} 
\centerline{Samuel Bosch,$^{1,3}$
Giacomo De Palma,$^{5}$, 
William M. Kaminsky,$^1$}
\centerline{Zi-Wen Liu,$^7$ 
Milad Marvian$^{8}$ }
\bigskip

\centerline{1. Turing, Inc.
2. Department of Mechanical Engineering, MIT. }
\centerline{ 3. Department of Electrical Engineering and Computer Science,
MIT. 4. Hitachi Laboratory, Cambridge UK.} 
\centerline{ 5. Scuola Normale Superiore, Pisa IT. 6. Department
of Physics, MIT. 7. Perimeter Institute CA.} 
\centerline{ 8. Department of Physics, Department of Electrical and Computer Engineering,
UNM.}
\centerline{$^*$ to whom correspondence should be addressed: slloyd@mit.edu}

\vskip 1cm
\noindent{\it Abstract:}  The quantum singular value transformation
is a powerful quantum algorithm that allows one to apply a polynomial
transformation to the singular values of a matrix that is embedded
as a block of a unitary transformation.   This paper shows how
to perform the quantum singular value transformation 
for a matrix that can be embedded as a block of a Hamiltonian.  The
transformation can be implemented in a purely Hamiltonian context by
the alternating application of Hamiltonians for chosen intervals: it
is an example of the Quantum Alternating Operator Ansatz (generalized
QAOA).  We also show how to use the Hamiltonian quantum singular value transformation to perform {\it inverse} block encoding
to implement a unitary of which a given Hamiltonian is a block.
Inverse block encoding leads to novel procedures for
matrix multiplication and for solving differential equations
on quantum information processors in a purely Hamiltonian fashion.  

\vfill\eject
The quantum singular value transformation (QSVT) [1] shows that if one can embed a matrix
$A$ in a block of a performable unitary transformation, one can also
implement a transformation with the same singular vectors as $A$, but 
whose singular values are polynomial functions of the singular values
of $A$.   The quantum singular value transformation 
is a powerful quantum algorithm
that encompasses other quantum algorithms such as quantum simulation
and matrix inversion as subcases.  The methods of the QSVT are 
based on the technique of qubitization [2], a state-of-the-art
method for simulating a Hamiltonian evolution given the ability
to perform block encoding.

This paper presents a purely Hamiltonian
version of the quantum singular value transformation.  
We show that if one has the ability to apply a Hamiltonian $H$ of which $A$ is
a block, then we can perform the full quantum singular value transformation 
using the Quantum Alternating Operator Ansatz (generalized QAOA):
we alternate application of the Hamiltonian $H$ with a second, 
readily applied Hamiltonian, for a chosen set of times.  
Our motivation in developing Hamiltonian QSVT is simple:
quantum information processing is generically implemented via the application
of semiclassical control fields, which apply time-dependent Hamiltonians
to the quantum information bearing degrees of freedom.   In the gate
model of quantum computing, these control fields are used to implement
unitary quantum logic gates, which are applied to the 
qubits/qudits/qumodes in the quantum computer.   But many more possible
Hamiltonians can be applied, including global Hamiltonians,
as in the Quantum Approximate Optimization Algorithm (original QAOA) [3].
In general, the ability to apply a set of Hamiltonians 
governed by the application of time varying control fields allows one
to apply any effective Hamiltonian in the algebra generated by the set.  
Different coherent quantum information processors, e.g., coherent diabatic
quantum annealers [4], allow the application of the Hamiltonian quantum
singular value transformation to the set of Hamiltonians that are `native'
to those processors -- i.e., those that can be efficiently applied.  
By introducing a purely Hamiltonian version of the quantum singular
value transformation that bypasses the gate model of quantum computing, 
we hope to expand the power of near term quantum information processors. 

\bigskip\noindent{\it Main result:}

In this paper we show that if one can apply a Hamiltonian of which a matrix
$A$ is an off-diagonal block, we can deterministically
implement the unitary transformation
$$U_f = i \pmatrix{ \sqrt{ I - f(A^\dagger) f(A)} & f(A^\dagger) \cr
f(A) & - \sqrt{ I - f(A) f(A^\dagger)} \cr}, \eqno(1) $$ 
where $f(A)$ is the matrix with the same singular vectors as $A$,
and singular values that are a function $f$ of the
singular values of $A$.  

$U_f$ is easily verified to be unitary as long as
$f(A^\dagger) f(A) \leq I$.  (For example, expand the square root
as a power series.)
Note that $U$ is anti-Hermitian as well as unitary,
and so has eigenvalues $\pm i$.  
Like the unitary QSVT, the Hamiltonian singular value transformation takes time
$O(\sigma_{min}^{-1} \log(1/\epsilon))$ to perform the transformation
to accuracy $\epsilon$, where $\sigma_{min}$ is the smallest singular
value of $A$.  

Setting $f(x) = x$ yields {\it inverse block
encoding}, or inverse qubitization [2]: 
the ability to apply a Hamiltonian of which a matrix
is a block allows one to apply a unitary of which the matrix is a block.
Below, we apply inverse block encoding to give Hamiltonian versions
of matrix multiplication and the quantum solution of differential equations.

\bigskip\noindent{\it Preliminaries:}

Assume the ability to apply a Hamiltonian $\pm H$, where
$$H = \pmatrix{ \bullet & A^\dagger \cr A & \bullet \cr}, \eqno(2)$$ 
and the on-diagonal blocks are arbitrary.  We assume that $A^\dagger A
\leq I$.
We also assume the ability to apply the Hamiltonian $\pm Z$, where
$$ 
Z \equiv \pmatrix{ I & 0 \cr 0 & - I \cr}.\eqno(3)$$
As shown in [5], by applying $\pm H$ in alternation with $\pm Z$, we can 
average out/refocus the on-diagonal terms of $H$ in time $O(1)$, so without 
loss of generality we take $H$ to be of the form
$$H = \pmatrix{ 0 & A^\dagger \cr A & 0 \cr}. \eqno(4)$$

Write $A = \sum_j \sigma_j
|\ell_j\rangle\langle r_j|$, where $\sigma_j$ are the singular values of
$A$ and $|r_j\rangle$, $|\ell_j\rangle$ are the corresponding right and left
singular vectors.  
The Hamiltonian $H$ acts on the Hilbert space
${\cal H}_R \oplus {\cal H}_L$ consisting of the direct sum of the
the Hilbert spaces for the right and left singular vectors.
The eigenvectors of $H$ take the form
$$H \pmatrix{ |r_j\rangle \cr \pm |\ell_j\rangle\cr}
 = \pm \sigma_j \pmatrix{ |r_j\rangle \cr \pm |\ell_j\rangle\cr}.\eqno(5)$$ 
That is, within the two-dimensional subspace ${\cal H}_j$ spanned by
$\pmatrix{ |r_j\rangle \cr \pm |\ell_j\rangle \cr}$, $H$ acts as $\sigma_j X$,
where $X$ is the $x$ Pauli matrix in this subspace.

Note that we can also apply the transformation
$$e^{i\phi Z /2} e^{-iHt} e^{-i\phi Z/2}
\equiv e^{-iG_\phi t},\eqno(6)$$
where
$$ G_\phi \equiv \pmatrix{ 0 & e^{i \phi} A^\dagger \cr  e^{-i\phi} A & 0 \cr}.
\eqno(7)$$
$G_\phi$ acts within ${\cal H}_j$ as $\sigma_j$  times
a  Pauli matrix for a rotation about an axis at angle
$\phi$ in the $xy$ plane.

\bigskip\noindent{\it Hamiltonian quantum singular value transformation:}

From the results of the previous section,
we can apply transformations of the form
$$e^{-i  G_{\phi_k} t_k}  \ldots 
 e^{-i  G_{\phi_1} t_1}.\eqno(8)$$
Equation (8) shows that we can generate unitary transformations that act
in each of the ${\cal H}_j$ as polynomials in $\cos (\sigma_j t_i)$,
$\sin(\sigma_j t_i)$.   The coefficents of these polynomials are 
determined by tuning the $\phi_i$.  Qubitization [2] shows how 
to generate a broad class of such polynomial unitary transformations. 
Note that because of the
Hamiltonian nature of our construction, we have greater flexibility than
the transformations applied in qubitization [2] and the quantum singular 
value transformation [1]: there, because $H$ is applied as a block of
a unitary, the times $t_i$ are all fixed to be equal.  Here, we can vary
the times $t_i$ as well as the angles $\phi_i$.  This feature could prove
useful for optimizing the transformation using variational methods.

The theorems of [1-2] show that the ability to perform transformations
of the form (8) allows us to 
apply transformations of the form
$$ U = \bigoplus_j \pmatrix{ P(\cos\sigma_j) & i\sin\sigma_j Q( \cos\sigma_j)\cr
i\sin\sigma_j Q^*( \cos\sigma_j) & P^*(\cos\sigma_j)}, \eqno(9)$$
where the degree of the polynomial $P$ is less than or equal to $k$,
and the degree of the polynomial $Q$ is less than or equal to $k-1$.
The polynomials $P$ and $Q$ can be determined at will by choosing
the angles $\phi_i$, subject to the constraint that $P$ has parity
$k$ mod 2, $Q$ has parity $k-1$ mod 2, and that
$|P(\cos \sigma)|^2 + \sin^2 \sigma |Q(\cos \sigma)|^2 = 1$.
The first of these constraints comes from the product form of equation (8),
and the second from unitarity.

To perform the Hamiltonian singular value transformation, we choose
$Q$ so that 
$$\sin\sigma_j Q^*( \cos\sigma_j) = f(\sigma_j) \pm \epsilon \eqno(10)$$
Equivalently, $\sqrt{1-x^2} Q^*(x) \approx f(\arccos(x))$: as in [1-2]
we are effectively decomposing the function $f(\arccos(x))$ in terms
of Chebyshev polynomials.   For suitably smooth functions $f$ the
decomposition achieves accuracy $\epsilon$ for Chebyshev polyomials
of degree $O(\log(1/\epsilon))$. 
Similarly, take
$$ -iP(\cos(\sigma_j)) = \sqrt{ I - f^*(\sigma_j) f(\sigma_j)}
\pm \epsilon.  \eqno(11)$$
To within accuracy $\epsilon$, the resulting unitary transformation is
$$U_f = i \pmatrix{ \sqrt{ I - f(A^\dagger) f(A)} & f(A^\dagger) \cr
f(A) & - \sqrt{ I - f(A) f(A^\dagger)} \cr}, \eqno(12) $$
which is the desired Hamiltonian singular value transformation of 
equation (1).

Inverse block encoding is obtained by taking $f(\sigma) = \sigma$,
which gives
$$U = i \pmatrix{ \sqrt{ I - A^\dagger A} & A^\dagger \cr
A & - \sqrt{ I - A A^\dagger}\cr}.	 \eqno(13) $$
Because the derivative of $\arccos$ diverges at $\pm 1$, we require
$|\sigma| < 1-\delta$ in order to obtain an error scaling of
$ \epsilon = O(e^{-\sqrt{2\delta} k})$ in the $k$-th order
Chebyshev polynomial expansion of $\arccos(x)$ [6-7]. 

The error scaling for the Hamiltonian quantum singular value transformation
is the same as for the unitary QSVT, and
the number of alternating steps required to perform the transformation
goes as $ \# = O( \sigma_{min}^{-1}\log(1/\epsilon))$.   
In comparison with performing the same procedure using the generalized
quantum linear systems algorithm [5,8], which relies on quantum phase
estimation, we see that the dependence of the time has gone
from $1/\epsilon$ to $\log(1/\epsilon)$. 

Since in practice the alternating Hamiltonians are applied by turning on and 
off semiclassical control fields, we also need to address the question of
how accurately these control fields need to be applied.   If each Hamiltonian
application consists of a time-dependent term of the form $\gamma(t) H$, then
the key figure of merit is the total integral of $\gamma(t)$ over the period
during which the control is turned on and off: this integral needs to
be precise to accuracy $O(\epsilon/ \#)$ in order for the accumulated error
over the overall alternating sequence to be less than $\epsilon$.   Note that
in existing high-coherence quantum information processors such as ion traps,
the semiclassical control error is the dominant source of error over the
quantum computation.    Current superconducting and ion-trap quantum computers
have pulse control errors on the order of $ 2*10^{-3} - 5*10^{-3}$, with a
next-generation goal of $1*10^{-3}$.   Given the favorable error
scaling, the Hamiltonian QSVT should be performable on current NISQ devices.

The Hamiltonian quantum singular value transformation is implemented solely through the alternating application of the Hamiltonian $H$ and the Hamiltonian $Z$.
That is, it is an application of the Quantum Alternating Operator Ansatz
or generalized QAOA.    As in the original QAOA (Quantum Approximate
Optimization Algorithm) [3], this feature makes the Hamiltonian QSVT ripe for
the application of variational methods: 
we set some task for the unitary
$U_f$ to accomplish and try to find an optimal $f$ by varying the parameters
$t_i,\phi_i$ in equation (8). 

\bigskip\noindent{\it Applications:}

Inverse block encoding provides a novel method for quantum
matrix multiplication [9-10]: from equation (13) we have
$$ U \pmatrix{ |\psi\rangle \cr 0 \cr} = i\pmatrix{ (\sqrt{I - A^\dagger A})
|\psi\rangle \cr A |\psi\rangle \cr},\eqno(14)$$
which allows one to implement $A|\psi\rangle$ with probability
$$\langle \psi | A^\dagger A |\psi\rangle.\eqno(15)$$  
Amplitude amplification provides a quadratic improvement, allowing the state
$A|\psi\rangle$ to be prepared in time 
$\sqrt{ \psi | A^\dagger A |\psi\rangle}$.

When $A$ is square, applying this procedure first to the
first and second components 
of the state $(|\psi\rangle, 0, \ldots, 0)^T$,
then to the second and third components, then to the third and fourth,
etc.
allows one deterministically to construct the state
$$\pmatrix{ (\sqrt{I - A^\dagger A}) |\psi\rangle
\cr
(\sqrt{I - A^\dagger A}) A |\psi\rangle \cr
\ldots \cr (\sqrt{I - A^\dagger A}) A^{n-1}|\psi\rangle \cr
A^n |\psi \rangle \cr} = 
\sum_{k=0}^{n-1} (\sqrt{I - A^\dagger A}) A^k |\psi\rangle|k\rangle 
+ A^n |\psi\rangle |n\rangle.\eqno(16)$$
Measuring the value of the second register yields the state
$ A^n |\psi\rangle$ with probability 
$$\langle \psi| {A^\dagger}^n A^n |\psi\rangle. \eqno(17)$$
Again, amplitude amplification provides a quadratic speed up in the
preparation of this state.  

Note that the requirement $A^\dagger A \leq I$ means that
the probability of obtaining $A^n|\psi\rangle$ will be
asymptotically exponentially suppressed except for components
corresponding to the singular value $\sigma_j = 1$.
Since $\sigma_j = 1$ corresponds to subspaces on which
$H$ acts in a unitary fashion, this method allows us to identify
`hidden' unitaries in $H$.

When $A = I + B\Delta t$, then the state (16) yields 
the Euler forward solution of the differential equation
$$ {d|\psi\rangle \over dt} = B |\psi\rangle. \eqno(18)$$
The requirement that 
$$A^\dagger A =  I + (B+B^\dagger) \Delta t +
B^\dagger B \Delta t^2 \leq I \eqno(18)$$ 
means that the eigenvalues of $B$ must have negative real part, so
that the differential equation (18) is dissipative.
If $B$ is sparse or low rank, then we can implement the Hamiltonian
$H$ via conventional methods of quantum Hamiltonian simulation [11-13].
This method of solving the forward differential equation allows
one to obtain the (unnormalized) state $|\psi(t)\rangle =
e^{Bt} |\psi(0)\rangle$ that is the solution to the differential
equation (18) by projecting onto the final component of the state
in (16), which
succeeds with probability $\langle \psi(t)|\psi(t)\rangle$.  

That is, one can use inverse block encoding to
solve dissipative differential equations on a quantum
computer without
having to resort to matrix inversion, a potential savings
in computational complexity
over conventional quantum differential equation solvers [14-15]. 

\bigskip\noindent{\it History states:}

The Hamiltonian version of the quantum singular 
value transformation 
can be used to perform quantum matrix inversion
to apply $(\sqrt{I - A^\dagger A})^{-1}$ to the first
$n$ components of the vector (16) to obtain the state
$$\pmatrix{ |\psi\rangle \cr A|\psi\rangle \cr \ldots \cr A^n |\psi\rangle}
= \sum_{k=0}^{n} A^k |\psi\rangle|k\rangle. \eqno(20) $$ 
In the case that $A = I + B\Delta t$, this state is the `history state'
for the solution of the differential equation (18).
The matrix inversion is non-deterministic, and succeeds with 
probability $O(1/\tilde\kappa)$, where $\tilde\kappa$ is the
condition number of $\sqrt{I - A^\dagger A}$.

\bigskip\noindent{\it Multiple Hamiltonians:}

When building
quantum computers, we implement unitary transformations by applying
time dependent Hamiltonians $\sum_j \gamma_j(t) H_j$
governed by semiclassical control fields $\gamma_j(t)$.  This ability
allows us to implement effective Hamiltonians in the Lie algebra
${\cal A}$ generated by the $H_j$, and to implement the corresponding
embedding unitaries.  The Hamiltonian QSVT then
allows us deterministically to implement the $U_f$ of equation (1)
for any $A$ that can be embedded in a block of a Hamiltonian
 $ \in {\cal A}$.   Note that even systems where the
applied Hamiltonians are quite simple, as in QAOA [3] where one alternates
the application of an Ising Hamiltonian with identical rotations
of the qubits about the $X$ axis, the algebra generated by
the Hamiltonians is typically the full algebra of all possible
Hamiltonians -- QAOA is universal for quantum computation [16].   
The set of effective Hamiltonians that can be generated {\it efficiently}
by any given coherently controllable quantum system depends
on the nature of the control Hamiltonians.  Since Hamiltonian
QSVT can be applied to any coherently controllable Hamiltonian
system, we hope that our results will encourage experimentalists to 
investigate novel architectures for Hamiltonian quantum information
processing.   For example, Hamiltonian QSVT is a natural procedure to
implement on the next generation of coherent diabatic quantum
annealers.

\bigskip\noindent{\it Discussion:}

This paper showed how to perform a Hamiltonian version of the quantum
singular value transformation solely by the alternating application of
Hamiltonians for different lengths of time.   The method is an  
example of the quantum alternating operator ansatz (generalized
QAOA), and allows the direct construction of the unitary transformations
of the QSVT, without the application of quantum logic gates.
Our goal in presenting this construction is straightforward: quantum
information processing is typically implemented by the application
of semiclassical control fields to implement time-varying Hamiltonians,
which can be applied globally as well as locally.    For experimentally
accessible Hamiltonians, the Hamiltonian QSVT represents a direct
and simple way to perform the powerful quantum singular value transformation.

As a simple example of Hamiltonian QSVT, we showed how to
perform inverse block encoding: given the
ability to apply a Hamiltonian, deterministically
implement a unitary transformation in which the Hamiltonian is
embedded as the off-diagonal blocks.   Inverse block encoding 
provides
novel methods for quantum matrix multiplication and the quantum
solution of forward differential equations.  Given the ability
to apply multiple Hamiltonians, inverse block encoding allows
one to embed any transformation in the algebra generated
by the set of accessible Hamiltonians in a deterministically
constructed unitary transformation.    

\vfill

\vfil\eject
\noindent{\it References}

\bigskip\noindent [1] 
A. Gily\'en, Y. Su, G.H. Low, N. Wiebe, {\it STOC 2019: Proceedings 
of the 51st Annual ACM SIGACT Symposium on Theory of Computing,}
193–204 (2019).  

\bigskip\noindent [2] G.H. Low, I.L. Chuang,
{\it Quantum} {\bf 3}, 163 (2019); arXiv: 1610.06546v3.

\bigskip\noindent [3] E. Farhi, J. Goldstone, S. Gutmann,
`A quantum approximate optimization algorithm,' arXiv: 1411.4028 (2014).

\bigskip\noindent [4] E.J. Crosson, D.A. Lidar, `Prospects for Quantum 
Enhancement with Diabatic Quantum Annealing,' arXiv: 2008.09913 (2020).

\bigskip\noindent [5] S. Lloyd, S. Bosch, G. De Palma, B. Kiani, Z.-W. Liu, 
M. Marvian, P. Rebentrost, D.M. Arvidsson-Shukur
`Quantum polar decomposition algorithm,' arXiv: 2006.00841 (2020).

\bigskip\noindent [6] J.P. Boyd, {\it Chebyshev and Fourier Spectral 
Methods,}, Dover Books on Mathematics, Dover (2001).

\bigskip\noindent [7] A. Gil, J. Segura, N.M. Temme, {\it Numerical Methods
for Special Functions,} Society of Industrial and Applied Mathematics (2007).

\bigskip\noindent [8] A.W. Harrow, A. Hassidim, S. Lloyd,
{\it Phys. Rev. Lett.} {\bf 103}, 150502 (2009).

\bigskip\noindent [9] C. Shao,
`Quantum Algorithms to Matrix Multiplication,' arXiv: 1803.01601.

\bigskip\noindent [10] L. Zhao, Z. Zhao, P. Rebentrost, J. Fitzsimons
`Compiling basic linear algebra subroutines for quantum computers,'
arXiv: 1902.10394.

\bigskip\noindent [11] S. Lloyd, {\it Science} {\bf 273},
1073-1078 (1996).

\bigskip\noindent [12] I. M. Georgescu, S. Ashhab, F. Nori,
{\it Rev. Mod. Phys.} {\bf 86}, 153 (2014).

\bigskip\noindent [13] A.M. Childs, A. Ostrander, Y. Su,
{\it Quantum} {\bf 3}, 182 (2019); arXiv: 1805.08385v2.

\bigskip\noindent [14] 
D.W. Berry, {\it J. Phys. A: Math. Theor.} {\bf 47} 105301 (2014).

\bigskip\noindent [15] 
D.W. Berry, A.M. Childs, A. Ostrander, G. Wang,
{\it Comm. Math. Phys.} {\bf 356}, 1057-1081 (2017). 

\bigskip\noindent [16] S. Lloyd, `Quantum approximate optimization
is computationally universal,' arXiv: 1812.11075 (2018). 

\vfill\eject\end